\documentclass[twocolumn,english,aps,prl,twocolumn,superscriptaddress]{revtex4}
\usepackage[T1]{fontenc}
\usepackage[latin9]{inputenc}
\usepackage{amsmath}
\usepackage{amssymb}
\usepackage{graphicx}
\usepackage{color}
\usepackage{bm}

\newcommand{\Tr}{\mathop{\text{Tr}}\nolimits}
\newcommand{\ket}[1]{|{#1}\rangle}
\newcommand{\bra}[1]{\langle{#1}|}

\newcommand{\ketbras}[3]{\ket{#1}_{#3}\hspace*{-0.2mm}\bra{#2}}

\makeatletter
\@ifundefined{textcolor}{}
{
 \definecolor{BLACK}{gray}{0}
 \definecolor{WHITE}{gray}{1}
 \definecolor{RED}{rgb}{1,0,0}
 \definecolor{GREEN}{rgb}{0,1,0}
 \definecolor{BLUE}{rgb}{0,0,1}
 \definecolor{CYAN}{cmyk}{1,0,0,0}
 \definecolor{MAGENTA}{cmyk}{0,1,0,0}
 \definecolor{YELLOW}{cmyk}{0,0,1,0}
}

\makeatother

\usepackage{babel}
\begin{document}

\title{
Quantum Computing in Plato's Cave\\
}

\author{Daniel Burgarth}
\affiliation{Department of Mathematics and Physics, Aberystwyth University, SY23
3BZ Aberystwyth, United Kingdom}

\author{Paolo Facchi}
\affiliation{Dipartimento di Fisica and MECENAS, Universit\`a di Bari, I-70126 Bari, Italy}
\affiliation{INFN, Sezione di Bari, I-70126 Bari, Italy}

\author{Vittorio Giovannetti}
\affiliation{NEST, Scuola Normale Superiore and Istituto Nanoscienze-CNR, I-56126 Pisa, Italy}

\author{Hiromichi Nakazato}
\affiliation{Department of Physics, Waseda University, Tokyo 169-8555, Japan}

\author{Saverio Pascazio}
\affiliation{Dipartimento di Fisica and MECENAS, Universit\`a di Bari, I-70126 Bari, Italy}
\affiliation{INFN, Sezione di Bari, I-70126 Bari, Italy}

\author{Kazuya Yuasa}
\affiliation{Department of Physics, Waseda University, Tokyo 169-8555, Japan}
\begin{abstract}
We show that mere observation of a quantum system can turn its dynamics from a very simple one into a universal quantum computation. This effect, which occurs if the system is regularly observed at short time intervals, can be rephrased as a modern version of Plato's Cave allegory.
More precisely, while in the original version of the myth, the reality perceived within the Cave is described by the projected shadows of some more fundamental dynamics which is intrinsically more complex, we found that in the quantum world the situation changes drastically as the `projected' reality perceived through sequences of measurements can be more complex than the one that originated it. After discussing examples we go on to show that this effect is generally to be expected: almost any quantum dynamics will become universal once `observed' as outlined above. Conversely, we show that any complex quantum dynamics can be `purified' into a simpler one in larger dimensions.
\end{abstract}

\maketitle

In the last 30 years the possibility of using quantum effects to develop an alternative approach to engineering has emerged as a realistic 
way to improve the efficiency of computation, communication and metrology \cite{qtech,qtech1,qtech11,qtech2}. 
At the very core of this revolutionary idea, the possibility of designing arbitrary dynamics of  quantum systems without spoiling the rather fragile correlations characterizing them is crucial.
What experimentalists typically do is to apply sequences of control pulses (e.g., by sequentially switching on and off different electromagnetic fields) to steer quantum systems.
In the quantum world, however, there is another option associated with the fact that the measurement process itself can induce a transformation on a quantum system.
In this context an intriguing possibility is offered by the quantum Zeno effect \cite{MS,ZenoMP}. 
It forces the system to evolve in a given subspace of the total Hilbert space
by performing frequent  projective measurements (Zeno dynamics) \cite{ZenoNatCommun,ZenoExpHaroche2014,ZenoSubspaces}, without the need of monitoring their outcomes (\emph{non-adaptive} feedback strategy). 
Several attempts have already been discussed  to exploit such effects for quantum computation, see e.g.,  \cite{ZCONTROL3,ref:FransonZeno,ref:ControlDecoZeno,ref:LeungRalphPRA,ref:LeungRalphNJP,ref:MyersGilchrist-ZenoPRA,ZCONTROL4,ZCONTROL1,ZCONTROL2,ZCONTROL5}.
In this work we show that the constraint imposed via a Zeno projection
   can in fact \emph{enrich} the dynamics  induced by a series of control pulses, allowing 
        the system of interest  to explore an algebra 
        that is  \emph{exponentially larger} than the original one. 
        In particular this effect can be used to 
 turn  a small set of quantum gates 
 into a universal set. 
Furthermore, exploiting the non-adaptive character of the scheme, we show that this Zeno enhancement  can also be implemented by a
 non-cooperative party, e.g.,  by noisy environment.

By the Zeno effect, the dynamics of the system is forced to evolve in a given subspace of the total Hilbert space \cite{ZenoNatCommun,ZenoExpHaroche2014,ZenoSubspaces}.
One might therefore think that the constrained dynamics is less ``rich'' than the original one. 
This naive expectation will turn out to be incorrect.
These surprising aspects of constraints bear interesting similarities to Einstein's precepts, according to which one can give a geometric description of complicated motion. 
The key geometrical idea is to embed the motion of the system of interest in a larger space, obtaining a forceless dynamics taking place along straight lines. The real 
dynamics, with interactions and potentials, is then obtained by projecting the system back onto the original 
space. 
Clearly, the constrained dynamics is more \emph{complex} than the higher-dimensional linear one. 
In classical mechanics these reduction procedures, linking a given dynamical system with the one constrained on a lower-dimensional manifold, have been extensively studied as an effective method for integrating the dynamics \cite{AM}.  
In particular, different classes of completely integrable systems arise as reductions of free ones with higher degrees of freedom \cite{KKS,OP,DAM}.
Notable examples include the three-dimensional Kepler problem, the Calogero-Moser model, Toda systems, KdV and other integrable systems.   
The moral is that in classical mechanics, by constraining the 
dynamics, one often obtains an increase in complexity.

Here we find a quantum version of this intriguing effect, which exploits the inherent non-commutative nature of quantum mechanics.
The main idea is that even if two Hamiltonians $H$ and $H'$ are commutative, their projected counterparts can be non-commutative
\begin{equation}
[H,H']=0
\quad
\not\Rightarrow\quad
[PHP,PH'P]=0,
\end{equation}
where $P=P^2$ is a projection.
Due to this fact we show that when 
passing from a set of control Hamiltonians $\{ H^{(1)}, \ldots, 
H^{(n)} \}$ to  
their 
projected versions  $\{ P {H}^{(1)} P,\ldots, P {H}^{(n)} P \}$ one can induce 
 an enhancement in the complexity of the system dynamics which can be exponential, to the extent that it can be used to transform a small number of quantum gates which are not universal 
into a universal set capable of performing arbitrary quantum-computational tasks. 
We find that this effect is completely general and happens in almost all systems.
Conversely, we prove that any complex dynamics can be viewed as a simple dynamics in a larger dimension, with the original dynamics realized as a projected dynamics.
What is interesting 
is that, in contrast to the classical case, the constraint which transforms a Hamiltonian $H$ into  $PHP$ can be imposed not by force but by  
a simple projective measurement whose outcomes need not be recorded (the process being effectively equivalent to the one associated with an external noise that is monitoring the system).

The underlying  mechanism can be rephrased as a modern version of Plato's Cave allegory \cite{Plato}. 
In the original version of the myth,  the reality perceived within the Cave is described by 
 the projected shadows of some more fundamental dynamics which is intrinsically more complex. In the quantum world, however, the situation changes drastically and the \emph{projected} reality perceived within Plato's Cave can be more complex than the one that has originated it.

\begin{figure}
\begin{center}
\includegraphics[width=\columnwidth]{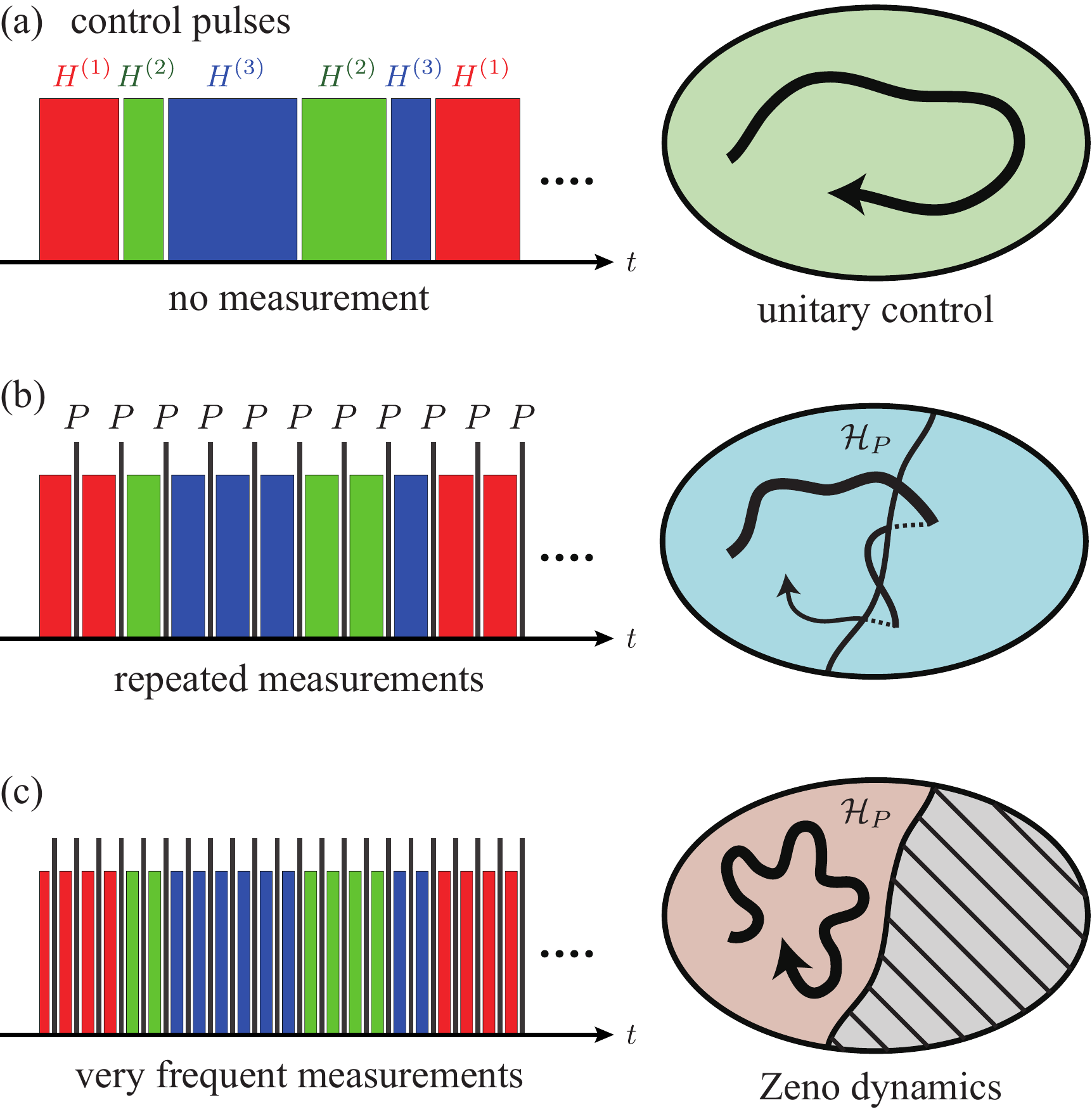}
\par\end{center}
\caption{
(a) We control a quantum system by switching on and off a set of given Hamiltonians $\{H^{(1)}, \ldots, H^{(n)}\}$.
(b) We perform projective measurements $P$ at regular time intervals during the control to check whether or not the state of the system belongs to a given subspace $\mathcal{H}_P$ of the global Hilbert space.
(c) In the limit of infinitely frequent measurements (Zeno limit), the system is confined in the subspace $\mathcal{H}_P$, where it evolves unitarily with the Zeno Hamiltonians $\{\bar{H}^{(1)}, \ldots, \bar{H}^{(n)}\}$ (Zeno dynamics).
The Zeno dynamics can explore the subspace $\mathcal{H}_P$ more thoroughly than the purely unitary control without measurement.
}
\label{fig:ZenoDynamics}
\end{figure}
\paragraph{Unitary control vs.\ Zeno dynamics.---} 
In controlled quantum dynamics, two Hamiltonians can commute, but their projected versions need not. 
This contains, in embryo, the simple idea discussed in the introductory paragraph:  interaction can arise from constraints (in this case projections).
To describe this mechanism it is worth reminding a few facts about the quantum control theory and the quantum Zeno effect.

In a typical quantum control scenario it is assumed that the system of interest (say the quantum register of a quantum computer, or the spins in an NMR experiment) can be externally driven by means of sequences of unitary 
pulses  $U^{(j)}= e^{-i H^{(j)}\tau}$,  activated by turning on and off a set of given Hamiltonians $\{H^{(1)}, \ldots, H^{(n)}\}$ [Fig.\ \ref{fig:ZenoDynamics}(a)] \cite{note:Traceless}.
If no limitations are imposed on the temporal durations $\tau$ of the pulses,
it is known \cite{DALE} that by properly arranging sequences composed of $\{U^{(1)},\ldots,U^{(n)}\}$ one can in fact force the system to evolve under the action of arbitrary transformations of the form  $U = e^{\Theta}$ 
with the anti-Hermitian operators $\Theta$ being elements of the real  Lie algebra $\mathfrak{L}=\mathfrak{Lie}(i H^{(1)}, \ldots, iH^{(n)})$
formed by the linear combinations of $i H^{(j)}$ and their iterated commutators, $[i H^{(j_1)},iH^{(j_2)}]$, $\bm{[}iH^{(j_1)},[iH^{(j_2)},iH^{(j_3)}]\bm{]}$, etc. 
Full controllability is hence achieved  if the dimension of $\mathfrak{L}$ is large enough to permit the implementation of all possible unitary transformations on the system, i.e.\ $\mathfrak{L}=\mathfrak{su}(d)$, with $d$ being the dimension of the system.

Suppose now that between the applications of consecutive pulses $U^{(j)}$ we are allowed to perform von Neumann's projective measurements 
[Fig.\ \ref{fig:ZenoDynamics}(b)], aimed at checking whether or not the state of the system belongs
to a given subspace $\mathcal{H}_P$ of the global Hilbert space.
Specifically, we will assume that the system is originally initialized in 
$\mathcal{H}_P$ while the various $U^{(j)}$ are infinitesimal transformations. 
Under this condition, the Zeno effect 
can be invoked, in the limit of infinitely frequent measurements, to ensure that with high probability
the system will be always found in $\mathcal{H}_P$ after each measurement, following a trajectory described by 
the  effective Hamiltonians $\bar{H}^{(j)} = PH^{(j)}P$, 
with $P$ the projection onto $\mathcal{H}_P$ [Fig.\ \ref{fig:ZenoDynamics}(c)] \cite{ZenoMP,ZenoSubspaces}.  
In other words, alternating the control pulses under the frequent applications of the projection $P$
the sequence $U^{(j_k)}  \cdots U^{(j_1)}$ can be effectively transformed into a rotation which on $\mathcal{H}_P$ is defined by the unitary operator
 $\bar{U}^{(j_k)}  \cdots  \bar{U}^{(j_1)}$
where $\bar{U}^{(j)} = e^{-i \bar{H}^{(j)} \tau}$.
Accordingly  the real  Lie algebra $\mathfrak{L}_\text{Zeno}=\mathfrak{Lie}(i \bar{H}^{(1)}, \ldots, i\bar{H}^{(n)})$ now replaces $\mathfrak{L}$
in defining the space of unitary transformations which can be forced upon the system. 
The fundamental result of this paper is to observe that 
by properly choosing the system setting, the dimension of $\mathfrak{L}_\text{Zeno}$
can be made larger than ${\mathfrak{L}}$,
 to the extent that 
the former can be used to fully control the system on $\mathcal{H}_P$, in spite of the fact that the latter 
is not capable of doing the same.
\begin{figure}
\begin{center}
\includegraphics[width=0.89\columnwidth]{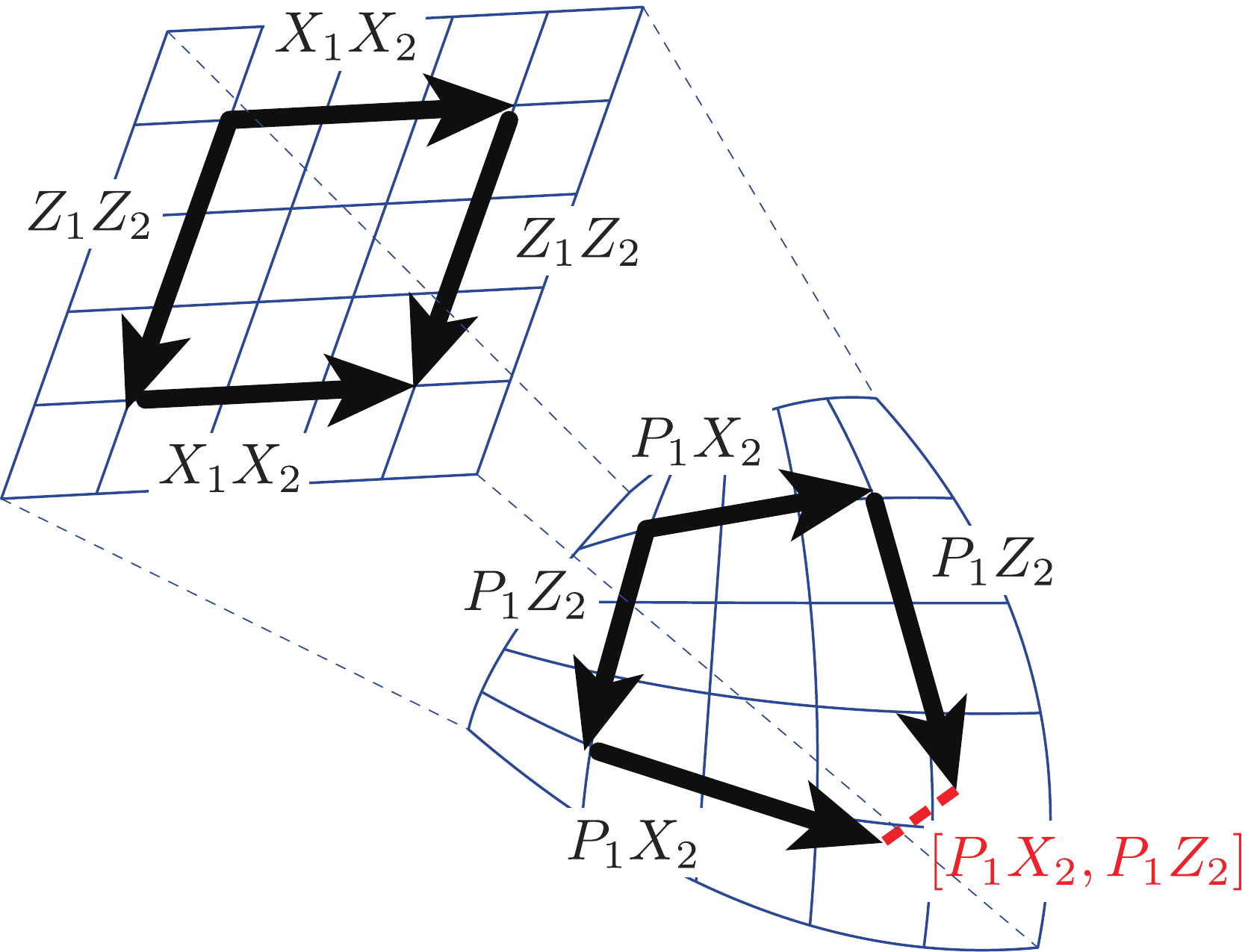}
\par\end{center}
\caption{
Schematics of the full versus projected system algebras. The arrows are tangents (generators) on a manifold of unitary transformations. In the larger space (upper), the operations commute, so no matter which way we go, we end up at the same point. 
It is not the case for the projected system (lower): the projected operations do not commute, and the gap represents the non-commutativity.
Even though the projected system is embedded in a smaller space, its dynamics is more complex, because of the curvature induced by the projection: new directions can be explored.}
\label{fig:Schematic-of-the}
\end{figure}

To better elucidate the idea we find it useful to introduce a simple example, where the system is identified with a  two-qubit system with control
 Hamiltonians 
\begin{equation}
H^{(1)}=
 X_{1}X_{2}
 ,\qquad
 H^{(2)}=
 Z_{1}Z_{2}, \label{controlH}
 \end{equation}
(we hereafter use  $X,Y,Z$ to denote Pauli operators, and write tensor products as strings, with systems being specified with subscripts and omitting the identity operators).
Notice that their commutator vanishes $[H^{(1)},H^{(2)}]=0$, and hence
the \emph{naked} algebra $\mathfrak{L}$ of the two-qubit system has dimension only 2.
Consider now the Zeno algebra induced by the projection 
\begin{equation} 
P_1=\frac{1+(X_{1}+Y_{1}+Z_{1})/\sqrt{3}}{2}
\equiv\ketbras{\phi}{\phi}{1},
\label{eqn:P1}
\end{equation}
which 
freezes the first qubit in the state $\ket{\phi}_1$ 
in the Zeno limit.
Then, the effective Zeno Hamiltonians  
\begin{eqnarray} 
&&\bar{H}^{(1)}=P_1{H}^{(1)}P_1=P_1X_{2}/\sqrt{3}, \nonumber \\
&&\bar{H}^{(2)}= P_1{H}^{(2)}P_1=P_1Z_{2}/\sqrt{3} \label{eq1}
\end{eqnarray}
exhibit a non-trivial commutator $[\bar{H}^{(1)},\bar{H}^{(2)}]=2iP_{1}Y_{2}/3$, which makes the dimension of   $\mathfrak{L}_\text{Zeno}$ equal to 3
(the situation is schematically illustrated  in Fig.\ \ref{fig:Schematic-of-the}). This in particular implies that  $\mathfrak{L}_\text{Zeno}$ 
can now be used to fully control the system in the subspace $\mathcal{H}_P=P_1(\mathbb{C}^2\otimes\mathbb{C}^2)$ (which is isomorphic to the Hilbert space of qubit 2), a task that could not be fulfilled with
the original ${\mathfrak{L}}$.

\paragraph{Zeno yields full control.---}
The example presented in the previous paragraph clarifies that
 the constrained dynamics can be more complex than the original unconstrained one.
The natural question arises: how big can such a difference become? To what extent can the presence of a measurement process increase the complexity of dynamics in quantum mechanics?
In the following we provide a couple of  examples  in which the enhancement in complexity is exponential.
While the unprojected dynamics are only two or three dimensional, the projected ones are \emph{univeral for quantum computation}. This shows that the simple ingredient of projective measurement can
strongly influence the complexity of dynamics.

\begin{figure}
\begin{center}
\includegraphics[width=0.8\columnwidth]{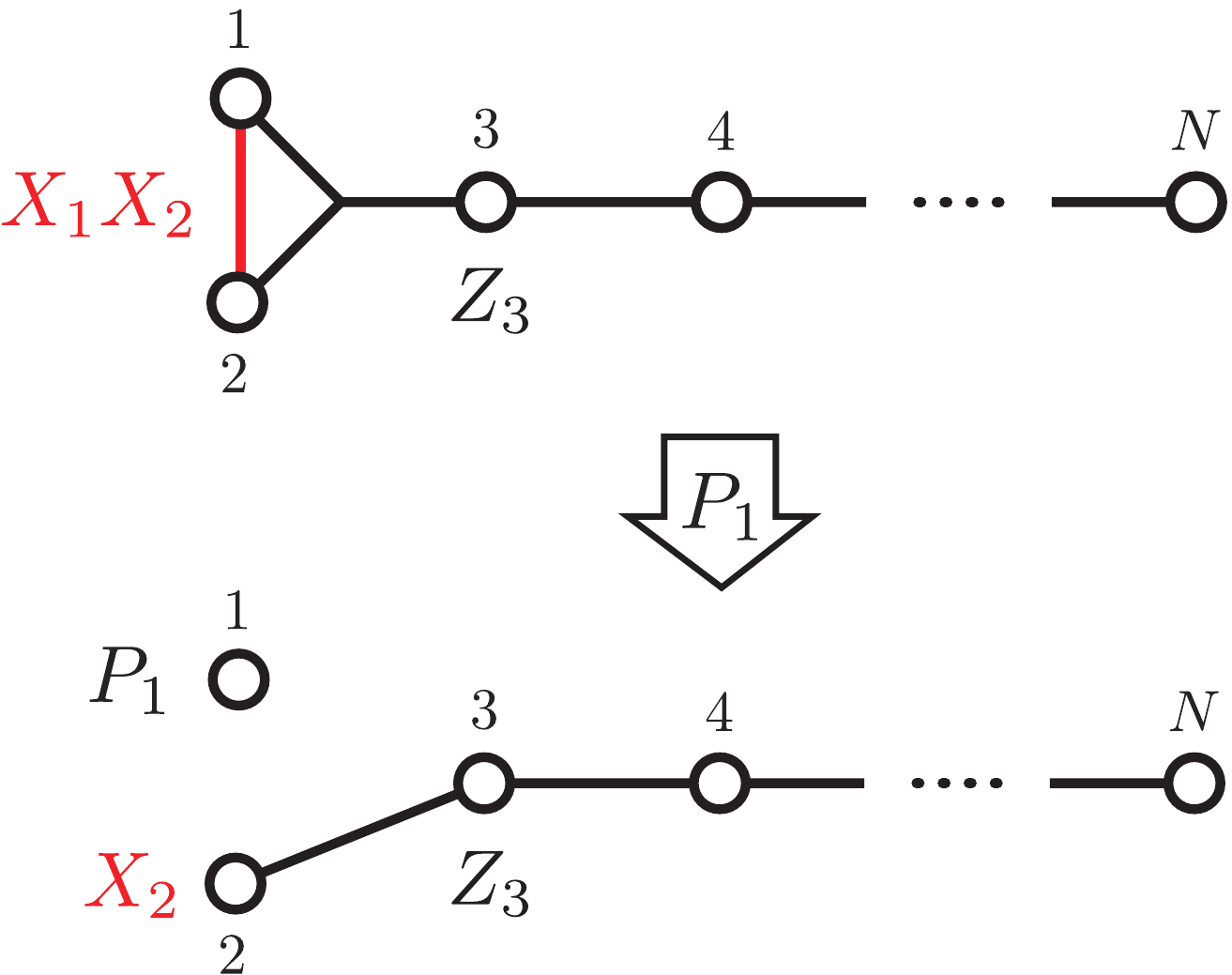}
\end{center}
\caption{
Schematics of the $N$-qubit model described in Example A of the text.
Straight edges represent the Heisenberg interactions, while the triple edge represents the three-body interaction among qubits 1--3.
The red part in the upper figure corresponds to $H^{(1)}$ acting on qubits 1 and 2, while the remainder including a local term $Z_3$ on qubit 3 corresponds to $H^{(2)}$ acting on all the $N$ qubits.
The Zeno projection $P_1$ on qubit 1 transforms the upper Hamiltonians to the lower model, where the state of  qubit 1 is frozen, while we are left with a Heisenberg chain with the local term $Z_3$ and a control $\bar{H}^{(2)}$ on qubit 2. The Lie algebra of the upper system is only two dimensional, while the lower allows us to perform full control over the system apart from the frozen qubit 1.}
\label{fig:model}
\end{figure}
Example A: Consider $N$ qubits (Fig.\ \ref{fig:model}, upper), the first two of which are manipulated via the control Hamiltonians $H^{(1)}=X_{1}X_{2}$,
and complement it with $H^{(2)}$  consisting of the nearest-neighbor Heisenberg interactions  
 ${H}^\text{(Heis)}_{3,\ldots,N} = \sum_{k=3}^{N-1}\left(XX+YY+ZZ\right)_{k,k+1}$ 
involving  all the qubits but the first two, together with a coupling term acting on the first three qubits and a local term on the third, i.e., 
\begin{equation} \label{THREE}
H^{(2)} = \sqrt{3}(X_{1}X_{2}X_{3}+Y_{1}Y_{2}Y_{3}+Z_{1}Z_{2}Z_{3}) 
+Z_3
+{H}^\text{(Heis)}_{3,\ldots,N}.
\end{equation}
Due to the anticommutation of the Pauli operators,  one can easily verify that the two Hamiltonians $H^{(1)}$ and  $H^{(2)}$ commute with each other $[H^{(1)},H^{(2)}]=0$, 
defining hence  a Lie algebra ${\mathfrak{L}}={\mathfrak{Lie}}( i H^{(1)}, iH^{(2)})$
which is barely two dimensional. 
Now let us consider their constrained versions using the same projection $P_{1}$ as in (\ref{eqn:P1}).
With this choice we have $\bar{H}^{(1)}=P_1{H}^{(1)}P_1=P_1X_{2}/\sqrt{3}$, and the Zeno Hamiltonian associated to ${H}^{(2)}$ is given by 
\begin{equation} \label{h3ham}
\bar{H}^{(2)} =P_{1}H^{(2)}P_{1}   
 =P_{1}(
 Z_3+ {H}^\text{(Heis)}_{2,\ldots,N}
 ), 
\end{equation}
where now ${H}^\text{(Heis)}_{2,\ldots,N}=\sum_{k=2}^{N-1}\left(XX+YY+ZZ\right)_{k,k+1}$ is  the nearest-neighbor Heisenberg Hamiltonian acting on qubits $2,\ldots,N$. 
While qubit 1 is kept frozen in the state $|\phi\rangle_1$ by the repetitive projections $P_1$, the remaining $N-1$ qubits now form a Heisenberg chain with a local term on qubit 3 (Fig.\ \ref{fig:model}, lower).
Elementary but cumbersome calculation shows that with these Zeno Hamiltonians qubit 2 is fully controllable, which by Ref.\ \cite{DB09} implies that the whole system apart from the frozen qubit 1 is fully controllable.
Consequently, we have  $\mathfrak{L}_\text{Zeno} = \mathfrak{Lie}(i \bar{H}^{(1)},i \bar{H}^{(2)})=P_{1}\mathop{\mathfrak{su}}(2^{N-1})$, so that the Zeno algebra is of exponential size, as claimed.

Example B: An alternative example which do not involve
 three-body interactions is available. Consider, for instance, three Hamiltonians $H^{(1)}=Z_{1}Z_{2}$, $H^{(2)}=X_{3}X_{4}$, and $H^{(3)}=\sqrt{3}H^\text{(Heis)}_{1,2} + \sqrt{3}H^\text{(Heis)}_{3,4}
+Z_{2}Z_{5}+Z_{5}+X_{4}X_{5}+X_{5}+H^\text{(Heis)}_{5,\ldots,N}$, and take the
Zeno projection to be  $P=P_1P_3$ with $P_1$ and $P_3$ projecting qubits 1 and 3 respectively into the states $\ket{\phi}_1$ and $\ket{\phi}_3$ defined as in (\ref{eqn:P1}).
These Hamiltonians commute with each other, and their Lie algebra $\mathfrak{L}={\mathfrak{Lie}}( i H^{(1)}, iH^{(2)}, iH^{(3)})$ is only three dimensional.
Analogously to the previous example, by exploiting the results of Ref.\ \cite{DB09} one can easily show that   the dimension of  ${\mathfrak{L}}_\text{Zeno}=\mathfrak{Lie}(i \bar{H}^{(1)}, i\bar{H}^{(2)}, i\bar{H}^{(3)})=P_1P_3\mathop{\mathfrak{su}}(2^{N-2})$ is again exponential, allowing the full control of all the qubits but the first and the third.

\paragraph{Generality and Hamiltonian purification.---}
What we have observed above is not a contrived phenomenon, but is actually a quite general one.
Considering the couple of Hamiltonians $H^{(1)}$ and $H^{(2)}$ with the projection $P_1$ of the above Example A, 
we are sure that there \emph{exist} a pair of commutative Hamiltonians and a projection such that the projected dynamics is essentially $\mathfrak{su}(2^{N-1})$. 
A standard argument in control theory is that if a system is fully controllable for a specific choice of parameters, then it is also fully controllable for almost all parameters \cite{DALE}. 
In our case it implies that \emph{almost all} commuting Hamiltonians will become universal through the Zeno projection on a single qubit (see Appendix for more details).

Furthermore, we can show the converse: any non-commutative dynamics can be thought of as the projected version of commutative dynamics in a larger space. This general phenomenon is in accord with the philosophy of geometrization discussed in the introduction.
In analogy with the purification of states in quantum information theory \cite{nielsen}, we call it \emph{Hamiltonian purification}. 
While we give a detailed mathematical analysis elsewhere, let us present the simplest case. 
Consider two arbitrary $d$-dimensional Hamiltonians $h^{(1)}$ and $h^{(2)}$.
We extend the Hilbert space by a single qubit and define their ``purifications'' by 
\begin{eqnarray}
&&H^{(1)}=1\otimes h^{(1)}+X\otimes h^{(2)},\nonumber\\
&&H^{(2)}=1\otimes h^{(2)}+X\otimes h^{(1)}.
\end{eqnarray}
These extended Hamiltonians $H^{(1)}$ and $H^{(2)}$ are easily seen to commute with each other, $[H^{(1)},H^{(2)}]=0$, and the projection by $P=(1+Z)\otimes 1/2$ yields $\bar{H}^{(1)}=PH^{(1)}P=(1+Z)\otimes h^{(1)}/2$ and $\bar{H}^{(2)}=PH^{(2)}P=(1+Z)\otimes h^{(2)}/2$, which act as $h^{(1)}$ and $h^{(2)}$ in the original space before the extension. 
We can furthermore apply this procedure iteratively to larger sets of Hamiltonians, which means that any complex dynamics can be thought of as a simple one taking place on a larger space, with the complexity arising only from projections. 

\paragraph{Local noise yields full control.---}
In a classical setting the measurement process is  typically perceived as a passive resource that enforces control only when 
properly inserted in a  feedback loop. As explicitly shown by our analysis, and more generally by 
the results of Refs.\ \cite{ZenoNatCommun,ZenoExpHaroche2014,ZenoSubspaces,ZCONTROL3,ref:FransonZeno,ref:ControlDecoZeno,ref:LeungRalphPRA,ref:LeungRalphNJP,ref:MyersGilchrist-ZenoPRA,ZCONTROL4,ZCONTROL1,ZCONTROL2,ZCONTROL5},
  this is no longer the case in quantum mechanics:  measurements
can indeed be used to directly  drive a quantum system even in the absence of a feedback mechanism.

Interestingly enough, for the control scheme we are analyzing here,
measurement is not the only way to implement the required projection $P$. 
The same effect is attainable by fast unitary kicks and by strong continuous coupling \cite{ZenoMP,ZenoSubspaces,FLP}.
Furthermore, owing to the non-adaptive character of the procedure (we never need to use the measurement outcomes to implement the control), it is also
achievable
by tailoring a strong dissipative process \cite{ref:VlatkoDissipAdiabat,ref:OreshkovCalsamiglia,ZCONTROL5,constrained_zeno}.
The latter option is of particular interest for us since, along the line of Refs.\ \cite{FV09,DIEHL,KRAUS},  it points out the possibility of taking advantages of the interaction of the system of interest with an external environment, which are  typically considered detrimental for quantum processing.

Specifically, for the qubit chain analyzed above (Example A), one can show  that the action of a simple amplitude damping channel \cite{nielsen} can raise the dynamical complexity to the level of universal quantum computation.
In fact, the decay process bringing qubit 1 to the state $\ket{\phi}_1$ can act as a projection $P_1$ (see Appendix), and in the strong-damping limit it is effective in inducing a quantum Zeno effect on qubit 1, yielding the full Lie algebra $\mathfrak{L}_\text{Zeno}$ in the rest of the qubit chain.
Moreover, due to the same reasoning as the one outlined above, almost all 
qubit amplitude damping channels induce exponential complexity.

\paragraph{Conclusions.---}
The schemes presented in this work 
are not meant to  be 
a  practical suggestion to implement quantum computers, because the implementation of a control scheme using Heisenberg chains would probably be inefficient (note however \cite{DB10}).
Instead they should be viewed as a proof of the fact that generally adding a simple projection or noise to a dynamical system can profoundly modify the global picture and provoke a drastic \emph{increase} in complexity.
This bears some similarities
to measurement-based quantum computation \cite{ref:MBQCPRL,ref:MBQCPRA}, although there are important differences, in that i) one does not require
the system to be initialized in a complex state, ii) the measurement is constant, and iii) its outcome is not used adaptively in future computations \cite{RJ13}.
 Our results can be presented as  a quantum version of the Plato's Cave myth, 
where the projection plays a more active role, making the dynamics of the associated \emph{quantum shadows} as complex as universal quantum computation; and, conversely through Hamiltonian purification, a non-commutative dynamics simple.

\paragraph{Acknowledgements.---}

This work was partially supported by PRIN 2010LLKJBX on ``Collective quantum phenomena: from strongly correlated systems to quantum simulators,'' by a Grant-in-Aid for Scientific Research, JSPS, by the Erasmus Mundus-BEAM Program, by a Grant for Excellent Graduate School from the Ministry of Education, Culture, Sports, Science and Technology (MEXT), Japan, and by a Waseda University Grant for Special Research Projects.

\section*{Appendix}
\paragraph{Sketch of the proof of the generality.---}
We found the two commuting Hamiltonians $H^{(1)}$ and $H^{(2)}$ in the $N$-qubit model depicted in Fig.\ \ref{fig:model} (Example A), whose projected counterparts $\bar{H}^{(1)}$ and $\bar{H}^{(2)}$ with the projection (\ref{eqn:P1}) of the structure $P=P_1\otimes1$ generate $\mathfrak{L}_\text{Zeno}=\mathfrak{Lie}(i \bar{H}^{(1)},i \bar{H}^{(2)})=P_{1}\mathop{\mathfrak{su}}(2^{N-1})$.
This single example makes us sure that it is the case for almost all systems.

To see this, let us formalize in the following way.
Take 
$(H^{(1)},H^{(2)},P)$ of Example A again.
We extract the relevant sector specified by $P$ from each element of $\mathfrak{L}_\text{Zeno}$ and call it $L_j$ ($j=1,\ldots,d^2-1$), which is a $d\times d$ matrix with dimension $d=2^{N-1}$ and is a function $L_j=L_j(H^{(1)},H^{(2)})$ of $H^{(1)}$ and $H^{(2)}$.
Together with the $d\times d$ identity matrix $L_0=1$, the matrices $\{L_j\}$ form $\mathfrak{u}(d)$.
This fact can be mathematically expressed as follows.
We ``vectorize'' each matrix $L_j$ to a $d^2$-dimensional column vector $|L_j)$ by lining up the columns of the matrix $L_j$ from top to bottom, and gather the column vectors $|L_j)$ side by side to make up a $d^2\times d^2$ matrix $\mathsf{L}=(|L_0)\,\ldots\,|L_{d^2-1}))$.
Then, the fact that the matrices $\{L_j\}$ span $\mathfrak{u}(d)$ is expressed as $D=\det\mathsf{L}\neq0$.
Note that this determinant is also a function $D=D(H^{(1)},H^{(2)})$ of $H^{(1)}$ and $H^{(2)}$.

Now take a generic couple of commuting Hamiltonians $\tilde{H}^{(1)}$ and $\tilde{H}^{(2)}$ of $N$ qubits, i.e., we randomly choose their eigenvalues $\{\tilde{\varepsilon}_1^{(1)},\ldots,\tilde{\varepsilon}_{2^N}^{(1)}\}$, $\{\tilde{\varepsilon}_1^{(2)},\ldots,\tilde{\varepsilon}_{2^N}^{(2)}\}$ and a common unitary matrix $\tilde{U}$ which diagonalizes $\tilde{H}^{(1)}$ and $\tilde{H}^{(2)}$ simultaneously.
Inserting this couple of Hamiltonians, the determinant $D(\tilde{H}^{(1)},\tilde{H}^{(2)})$ is, by construction, a polynomial in the parameters $\{\tilde{\varepsilon}_j^{(i)}, \tilde{U}_{kl}\}$ ($i=1,2; j,k,l=1,\ldots,2^N$).
We already know that this polynomial is non-vanishing for the parameter set $\{\varepsilon_j^{(i)}, U_{kl}\}$ corresponding to the above specific choice of the Hamiltonians $H^{(1)}$ and $H^{(2)}$.
Therefore, the determinant $D$ is a non-zero polynomial in the parameters $\{\tilde{\varepsilon}_j^{(i)}, \tilde{U}_{kl}\}$, implying that its roots are of measure zero in the parameter space.
In other words, for almost all parameters $\{\tilde{\varepsilon}_j^{(i)}, \tilde{U}_{kl}\}$, the determinant $D$ is non-vanishing, and in turn, almost all couples of commuting Hamiltonians become universal, generating $\mathfrak{L}_\text{Zeno}=P_{1}\mathop{\mathfrak{su}}(2^{N-1})$, by the projection $P$ on the first qubit.
This argument can be generalized to any rank $2^{N-1}$ projection, and also to any qubit amplitude damping channel in the strong-damping limit.

\paragraph{Projection by amplitude damping channel.---}
The continuous projection $P_1$ required for the qubit-chain model depicted in Fig.\ \ref{fig:model} can be induced by an amplitude damping channel acting on qubit 1.
In fact, consider the master equation 
$\dot{\rho}(t)
=-\frac{1}{2}\gamma(L^\dag L \rho + \rho L^\dag L-2L \rho L^\dag)$
with a single Lindblad operator $L= \ketbras{\phi}{\phi_\perp}{1}$ which
 describes the decay of qubit 1 from $\ket{\phi_\perp}_1$ to $\ket{\phi}_1$, where $\ket{\phi}_1$ is associated with the projection $P_1$ in (\ref{eqn:P1}) 
and $\ket{\phi_\perp}_1$ is the state orthogonal to $\ket{\phi}_1$. Solving the system dynamics under the 
master equation yields 
$\rho(t) = (1-e^{-\gamma t}) P_1\Tr_1\rho(0) 
+ e^{-\gamma t}[P_1\rho(0)P_1+Q_1 \rho(0) Q_1]
+ e^{-\gamma t/2}[P_1\rho(0)Q_1+Q_1 \rho(0) P_1]$, 
where $Q_1=1-P_1$, and $\Tr_1$ represents the partial trace over qubit 1.
Thus, in the limit $\gamma t \to \infty$, we have
$\rho(t)\to P_1\Tr_1\rho(0)$,
and qubit 1 is projected into the state $\ket{\phi}_1$ with probability 1.
If this process takes place on a time scale $\gamma^{-1}$ much shorter than any other time scales involved in the dynamics or the controls, then it is effective in inducing a quantum Zeno effect on qubit 1, and it is essentially equivalent to repeating projective measurements.


\begin{thebibliography}{38}
\expandafter\ifx\csname natexlab\endcsname\relax\def\natexlab#1{#1}\fi
\providecommand{\enquote}[1]{#1}

\bibitem[Dowling and Milburn(2003)]{qtech}
J.~P. Dowling and G.~J. Milburn, \enquote{Quantum technology: the second
  quantum revolution.} \emph{Phil. Trans. R. Soc. A} \textbf{361}, 1655--1674
  (2003).

\bibitem[Deutsch(2003)]{qtech1}
D.~Deutsch, \enquote{Physics, philosophy and quantum technology.} in
  \emph{Proceedings of the Sixth International Conference on Quantum
  Communication, Measurement and Computing}, edited by J.~H. Shapiro and
  O.~Hirota
 (Rinton Press, Princeton, NJ, 2003).

\bibitem[Zoller et~al.(2005)]{qtech11}
P.~Zoller, {Th. Beth}, D.~Binosi, R.~Blatt, H.~Briegel, D.~Bruss, T.~Calarco,
  J.~I. Cirac, D.~Deutsch, J.~Eisert, A.~Ekert, C.~Fabre, N.~Gisin,
  P.~Grangiere, M.~Grassl, S.~Haroche, A.~Imamoglu, A.~Karlson, J.~Kempe,
  L.~Kouwenhoven, S.~Kr\"oll, G.~Leuchs, M.~Lewenstein, D.~Loss,
  N.~L\"utkenhaus, S.~Massar, J.~E. Mooij, M.~B. Plenio, E.~Polzik, S.~Popescu,
  G.~Rempe, A.~Sergienko, D.~Suter, J.~Twamley, G.~Wendin, R.~Werner,
  A.~Winter, J.~Wrachtrup, and A.~Zeilinger, \enquote{Quantum information
  processing and communication.} \emph{Eur. Phys. J. D} \textbf{36}, 203--228
  (2005).

\bibitem[Kimble(2008)]{qtech2}
H.~J. Kimble, \enquote{The quantum internet.} \emph{Nature (London)}
  \textbf{453}, 1023--1030 (2008).

\bibitem[Misra and Sudarshan(1977)]{MS}
B.~Misra and E.~C.~G. Sudarshan, \enquote{The Zeno's paradox in quantum
  theory.} \emph{J. Math. Phys.} \textbf{18}, 756--763 (1977).

\bibitem[Facchi and Pascazio(2008)]{ZenoMP}
P.~Facchi and S.~Pascazio, \enquote{Quantum Zeno dynamics: mathematical and
  physical aspects.} \emph{J. Phys. A: Math. Theor.} \textbf{41}, 493001
  (2008).

\bibitem[Sch\"afer et~al.(2014)]{ZenoNatCommun}
F.~Sch\"afer, I.~Herrera, S.~Cherukattil, C.~Lovecchio, F.~S. Cataliotti,
  F.~Caruso, and A.~Smerzi, \enquote{Experimental realization of quantum zeno
  dynamics.} \emph{Nat. Commun.} \textbf{5}, 3194 (2014).

\bibitem[Signoles et~al.(2014)]{ZenoExpHaroche2014}
A.~Signoles, A.~Facon, D.~Grosso, I.~Dotsenko, S.~Haroche, J.-M. Raimond,
  M.~Brune, and S.~Gleyzes, \enquote{Confined quantum Zeno dynamics of a
  watched atomic arrow.} \emph{arXiv:1402.0111 [quant-ph]}  (2014).

\bibitem[Facchi and Pascazio(2002)]{ZenoSubspaces}
P.~Facchi and S.~Pascazio, \enquote{Quantum Zeno Subspaces.} \emph{Phys. Rev.
  Lett.} \textbf{89}, 080401 (2002).

\bibitem[Childs et~al.(2002)]{ZCONTROL3}
A.~M. Childs, E.~Deotto, E.~Farhi, J.~Goldstone, S.~Gutmann, and A.~J. Landahl,
  \enquote{Quantum search by measurement.} \emph{Phys. Rev. A} \textbf{66},
  032314 (2002).

\bibitem[Franson et~al.(2004)]{ref:FransonZeno}
J.~D. Franson, B.~C. Jacobs, and T.~B. Pittman, \enquote{Quantum computing
  using single photons and the Zeno effect.} \emph{Phys. Rev. A} \textbf{70},
  062302 (2004).

\bibitem[Facchi et~al.(2005)]{ref:ControlDecoZeno}
P.~Facchi, S.~Tasaki, S.~Pascazio, H.~Nakazato, A.~Tokuse, and D.~A. Lidar,
  \enquote{Control of decoherence: Analysis and comparison of three different
  strategies.} \emph{Phys. Rev. A} \textbf{71}, 022302 (2005).

\bibitem[Leung and Ralph(2006)]{ref:LeungRalphPRA}
P.~M. Leung and T.~C. Ralph, \enquote{Improving the fidelity of optical Zeno
  gates via distillation.} \emph{Phys. Rev. A} \textbf{74}, 062325 (2006).

\bibitem[Leung and Ralph(2007)]{ref:LeungRalphNJP}
P.~M. Leung and T.~C. Ralph, \enquote{Optical zeno gate: bounds for fault
  tolerant operation.} \emph{New J. Phys.} \textbf{9}, 224 (2007).

\bibitem[Myers and Gilchrist(2007)]{ref:MyersGilchrist-ZenoPRA}
C.~R. Myers and A.~Gilchrist, \enquote{Photon-loss-tolerant Zeno
  controlled-\textsc{sign} gate.} \emph{Phys. Rev. A} \textbf{75}, 052339
  (2007).

\bibitem[Aharonov and {Ta-Shma}(2007)]{ZCONTROL4}
D.~Aharonov and A.~{Ta-Shma}, \enquote{Adiabatic quantum state generation.}
  \emph{SIAM J. Comput.} \textbf{37}, 47--82 (2007).

\bibitem[Paz-Silva et~al.(2012)]{ZCONTROL1}
G.~A. Paz-Silva, A.~T. Rezakhani, J.~M. Dominy, and D.~A. Lidar, \enquote{Zeno
  effect for quantum computation and control.} \emph{Phys. Rev. Lett.}
  \textbf{108}, 080501 (2012).

\bibitem[Dominy et~al.(2013)]{ZCONTROL2}
J.~M. Dominy, G.~A. Paz-Silva, A.~T. Rezakhani, and D.~A. Lidar,
  \enquote{Analysis of the quantum Zeno effect for quantum control and
  computation.} \emph{J. Phys. A: Math. Theor.} \textbf{46}, 075306 (2013).

\bibitem[Burgarth et~al.(2013)]{ZCONTROL5}
D.~Burgarth, P.~Facchi, V.~Giovannetti, H.~Nakazato, S.~Pascazio, and K.~Yuasa,
  \enquote{Non-Abelian phases from quantum Zeno dynamics.} \emph{Phys. Rev. A}
  \textbf{88}, 042107 (2013).

\bibitem[Abraham and Marsden(1978)]{AM}
R.~Abraham and J.~E. Marsden, \emph{Foundations of Mechanics}
 (Westview Press, Cambridge, MA, 1978), 2nd edn.

\bibitem[Kazhdan et~al.(1978)]{KKS}
D.~Kazhdan, B.~Kostant, and S.~Sternberg, \enquote{Hamiltonian group actions
  and dynamical systems of calogero type.} \emph{Commun. Pure Appl. Math.}
  \textbf{31}, 481--507 (1978).

\bibitem[Olshanetsky and Perelomov(1981)]{OP}
M.~Olshanetsky and A.~Perelomov, \enquote{Classical integrable
  finite-dimensional systems related to Lie algebras.} \emph{Phys. Rep.}
  \textbf{71}, 313--400 (1981).

\bibitem[D'Avanzo and Marmo(2005)]{DAM}
A.~D'Avanzo and G.~Marmo, \enquote{Reduction and unfolding: the Kepler
  problem.} \emph{Int. J. Geom. Methods Mod. Phys.} \textbf{2}, 83--109 (2005).
  
\bibitem{Plato}
 Plato, \emph{The Republic.} (Oxford University Press, Oxford, 1941).

\bibitem[not(????)]{note:Traceless}
Without loss of generality the Hamiltonians can be assumed to be traceless,
  since the global phase does not play any role.

\bibitem[D'Alessandro(2008)]{DALE}
D.~D'Alessandro, \emph{Introduction to Quantum Control and Dynamics}
 (Champman \& Hall/CRC, Boca Raton, FL, 2008).

\bibitem[Burgarth et~al.(2009)]{DB09}
D.~Burgarth, S.~Bose, C.~Bruder, and V.~Giovannetti, \enquote{Local
  controllability of quantum networks.} \emph{Phys. Rev. A} \textbf{79},
  060305(R) (2009).

\bibitem[Nielsen and Chuang(2000)]{nielsen}
M.~A. Nielsen and I.~L. Chuang, \emph{Quantum Computation and Quantum
  Information}
 (Cambridge University Press, Cambridge, 2000).

\bibitem[Facchi et~al.(2004)]{FLP}
P.~Facchi, D.~A. Lidar, and S.~Pascazio, \enquote{Unification of dynamical
  decoupling and the quantum Zeno effect.} \emph{Phys. Rev. A} \textbf{69},
  032314 (2004).

\bibitem[Carollo et~al.(2006)]{ref:VlatkoDissipAdiabat}
A.~Carollo, M.~F. Santos, and V.~Vedral, \enquote{Coherent quantum evolution
  via reservoir driven holonomies.} \emph{Phys. Rev. Lett.} \textbf{96}, 020403
  (2006).

\bibitem[Oreshkov and Calsamiglia(2010)]{ref:OreshkovCalsamiglia}
O.~Oreshkov and J.~Calsamiglia, \enquote{Adiabatic Markovian dynamics.}
  \emph{Phys. Rev. Lett.} \textbf{105}, 050503 (2010).

\bibitem[Stannigel et~al.(2013)]{constrained_zeno}
K.~Stannigel, P.~Hauke, D.~Marcos, M.~Hafezi, S.~Diehl, M.~Dalmonte, and
  P.~Zoller, \enquote{Constrained dynamics via the Zeno effect in quantum
  simulation: implementing non-Abelian lattice gauge theories with cold atoms.}
  \emph{arXiv:1308.0528 [quant-ph]}  (2013).

\bibitem[Verstraete et~al.(2009)]{FV09}
F.~Verstraete, M.~M. Wolf, and J.~I. Cirac, \enquote{Quantum computation and
  quantum-state engineering driven by dissipation.} \emph{Nat. Phys.}
  \textbf{5}, 633--636 (2009).

\bibitem[Diehl et~al.(2008)]{DIEHL}
S.~Diehl, A.~Micheli, A.~Kantian, B.~Kraus, H.~P. B\"uchler, and P.~Zoller,
  \enquote{Quantum states and phases in driven open quantum systems with cold
  atoms.} \emph{Nat. Phys.} \textbf{4}, 878--883 (2008).

\bibitem[Kraus et~al.(2008)]{KRAUS}
B.~Kraus, H.~P. B\"uchler, S.~Diehl, A.~Kantian, A.~Micheli, and P.~Zoller,
  \enquote{Preparation of entangled states by quantum Markov processes.}
  \emph{Phys. Rev. A} \textbf{78}, 042307 (2008).

\bibitem[Burgarth et~al.(2010)]{DB10}
D.~Burgarth, K.~Maruyama, M.~Murphy, S.~Montangero, T.~Calarco, F.~Nori, and
  M.~B. Plenio, \enquote{Scalable quantum computation via local control of only
  two qubits.} \emph{Phys. Rev. A} \textbf{81}, 040303(R) (2010).

\bibitem[Raussendorf and Briegel(2001)]{ref:MBQCPRL}
R.~Raussendorf and H.~J. Briegel, \enquote{A one-way quantum computer.}
  \emph{Phys. Rev. Lett.} \textbf{86}, 5188--5191 (2001).

\bibitem[Raussendorf et~al.(2003)]{ref:MBQCPRA}
R.~Raussendorf, D.~E. Browne, and H.~J. Briegel, \enquote{Measurement-based
  quantum computation on cluster states.} \emph{Phys. Rev. A} \textbf{68},
  022312 (2003).

\bibitem[Jozsa and Van~den Nest(2013)]{RJ13}
R.~Jozsa and M.~Van~den Nest, \enquote{Classical simulation complexity of
  extended Clifford circuits.} \emph{arXiv:1305.6190 [quant-ph]}  (2013).

\end{thebibliography}
\end{document}